\documentclass[10pt, a4paper,onecolumn]{article} 
\newcommand{\bra}[1]{\langle #1|}
\newcommand{\ket}[1]{|#1\rangle}
\newcommand{\expt}[1]{\langle#1\rangle}
\newcommand{\braket}[2]{\langle #1|#2\rangle}
\title{Entropic Dynamics: A Hybrid-Contextual model of Quantum Mechanics}
\author{Kevin Vanslette\\ kvanslette@albany.edu\\
\\Department of Physics, University at Albany (SUNY)\\
Albany, NY 12222, USA}

\date{\today} 

\begin{document}

\maketitle 



\begin{abstract}
The Bell-KS theorem and the more recent $\psi$-epistemic \emph{no-go} theorems of QM are discussed in the context of Entropic Dynamics. In doing so we find that the Bell-KS theorem allows for, a perhaps overlooked, hybrid-contextual model of QM in which one set of commuting observables (position in this case) is non-contextual and all other observables are contextual. Entropic Dynamics is in a unique position as compared to other foundational theories of QM because it derives QM using standard techniques in Bayesian probability theory. In this formalism, position is the preferred basis from which inferences about other contextual operators are made. This leads to the interpretation that Entropic Dynamics is a hybrid-contextual model of QM, which we show to be consistent with the Bell-KS theorem and QM.
\end{abstract}


\section{Introduction}

Quantum Mechanics (QM) is an odd mix of the predictable and unpredictable. On one hand, it is hugely successful in its ability to predict the set of eigenvalues, expectation values, and operators for a particle-system of interest.  On the other hand, each measurement holds some amount of unpredictability, quantified by a probability distribution, except for a few trivial cases. This unpredictable nature leaves a space for the many interpretations of QM to coexist inharmoniously within the community - a community, no doubt, easily bothered by disharmony of any-type. 

The community reduces and organizes this disharmony by ruling out interpretations and foundational theories of QM that disagree with the predictable findings of QM. This is done by first making a few reasonable assumptions a theory of QM may obey, and then by showing these assumptions lead to contradictions in the formalism, construct a \emph{no-go} theorem. This is the basis of the Bell inequalities \cite{BellEPR}, the Bell-KS theorem \cite{Bell,KS,Mermin}, as well as the findings of Pusey-Barret-Rudolph (PBR) \cite{Pusey} (reviewed in \cite{Leifer}) on the epistemic interpretation of the wavefunction. Post-analysis, the final results are sometimes tabulated in 2 by 2 tables -- for example: 
\begin{center}
\begin{tabular}{c|c|c| }

   & $\psi$-ontic & $\psi$-epistemic \\ 
\hline
 contextual & A & B \\  
\hline
 noncontextual & C& D\\
\hline
\end{tabular}
,
\end{center}
and are followed by statements like, ``theories of type ``C" or ``D" are ruled out by the Bell-KS theorem and ``B" is ruled out by PBR". A reader may be inclined to conclude that QM must be a theory of type ``A" (potential interpretation of Bohmian mechanics).  The 2 by 2 is by nature an over simplification; it fails to span the entire set of plausible theories, and consequently interpretations, of QM. This is due the fact that $\emph{no-go}$ theorems are proofs by contradiction, and only theories which strictly adhere to their (shown invalid) assumptions are ruled out.

In particular, this paper will show Entropic Dynamics (ED) to be a theory of QM which lies on the line between theories ``B" and ``D", while not being ruled out by any of the aforementioned discussed \emph{no-go} theorems. We classify ED as a hybrid-contextual theory of QM because positions are treated noncontextually and all other observables are treated contextually -- the main result of this paper.  In the same way, other theories and interpretations of QM may slip between the cracks of these \emph{no-go} theorems, which reopens the perhaps presumed closed universe of discourse for a few edge theories and interpretations of QM.

ED will be reviewed briefly as it pertains to the \emph{no-go} theorems of interest. New insights of how these \emph{no-go} theorems are handled within ED will be presented and some critiques will be given. The sense in which a theory can be hybrid-contextual and still obey QM will be discussed.

\section{Entropic Dynamics}
 Entropic Dynamics is an application of inference and probabilistic techniques with the goal of extracting laws of physics. Here we are interested in the constraints and assumptions required to derive Quantum Mechanics from the first principles of inference and probability updating \cite{ED, book, Jaynes1, Jaynes2,Jaynesbook,Cox}. Here we briefly review the aspects of ED pertaining to the \emph{no-go} theorems of interest. An extended review can be found at \cite{ED}.  

The first step in any inferences problem is to state the universe of discourse, the set of possible outcomes or microstates, one would like to infer on the basis of incomplete information. To derive Quantum Mechanics (in flat space) the universe of discourse spanned by $N$ particles are their positions in a flat Euclidean space $\mathbf{X}$ (metric $\delta _{ab}$). Our knowledge of the positions of particles is characterized by a probability density $\rho(x)$ where $x$ is a coordinate in a $3N$ dimensional configuration space of particle coordinates $x^a_n$, where $a=1,2,3$ denotes the $a$th spacial axis of the $n$th particle's position. When convenient we may use a super-index notation $x^A\equiv x^a_n$ where $A=(n,a)$. From the onset, particles have \emph{definite} yet \emph{unknown} positions and are treated as the ``physical" or ``ontological" quantities we are interested in inferring. QM will be derived using this universe of discourse -- namely that particle positions are noncontextual.

Now that the microstates have been specified, we are inclined to ask how the position of these particles change. In particular, if the particles are located at $x$, we wish to know how probable it is for $x\rightarrow x'$, that is, we seek a probability of the form $P(x'|x)$ to quantify this uncertainty. Not knowing anything about how particles change position gives trivial dynamics, so we make the following assumptions: 1) particles move along continuous trajectories and 2) particles may have a tendency to be correlated and drift. These assumptions are represented by expectation value constraints on $P(x'|x)$. The first assumption is saturated by making large $\Delta x^a_n=x^{'a}_n-x^a_n$ improbable. This is done by imposing $P(x'|x)$ have small variances, $\kappa_n$, in particle coordinates,  
\begin{equation}
\langle \Delta x_{n}^{a}\Delta x_{n}^{b}\rangle \delta _{ab}=\kappa
_{n},\qquad (n=1,\ldots, N)~~  \label{kappa n}
\end{equation}%
(where motion is continuous in the limit $\kappa _{n}\rightarrow 0$). The second assumption is imposed by one additional constraint, 
\begin{equation}
\langle \Delta x^{A}\rangle \partial _{A}\phi
=\sum\limits_{n,a}\left\langle \Delta x_{n}^{a}\right\rangle \frac{%
\partial \phi(x) }{\partial x_{n}^{a}}=\kappa ^{\prime },  \label{kappa prime}
\end{equation}%
where $\kappa ^{\prime }$ is another small constant and $\frac{%
\partial \phi(x) }{\partial x_{n}^{a}}$ is the
\textquotedblleft drift\textquotedblright\ gradient. There are many probability distributions $P(x'|x)$ which satisfy the above expectation value constraints. We therefore use the method of maximum entropy \cite{book, Jaynes1, Jaynes2} to find the ``least biased" distribution by maximizing the entropy,
\begin{eqnarray}
S[P,Q]=-\int dx' P(x'|x)\log\frac{P(x'|x)}{Q(x'|x)},\label{entropy}
\end{eqnarray}
with respect to the expectation value constraints via the Lagrange multiplier method. Let $\{\alpha_n\}$ be the Lagrange multipliers which impose the $N$ constraints from (\ref{kappa n}) and let $\alpha'$ be the Lagrange multiplier which imposes (\ref{kappa prime}). Maximizing the above entropy with respect to these constraints and normalization gives the following probability distribution after completing the square,
\begin{eqnarray}
P(x'|x)=\frac{1}{Z}\exp\Big[-\frac{1}{2}\sum_n\alpha_n\delta_{ab}(\bigtriangleup x^a_n-{\bigtriangleup x^a_n})(\bigtriangleup x^b_n-{\bigtriangleup x^b_n})\Big],\label{4}
\end{eqnarray}
where $\expt{\Delta x^A}=\frac{\alpha'}{\alpha_n}\delta^{ab}\frac{\partial \phi}{\partial x^b_n}$. The prior distribution $Q(x'|x)$ is a very broad normalizable Gaussian distribution encoding that before constraints are made, particles may jump anywhere with near to equal probability, and so, it has been absorbed into the normalization constant. There are explicit and in-depth arguments for the notion of an instant in ED \cite{ED,book} that we need not explore here; however, in summary, $t$ is introduced as a convenient bookkeeping parameter to label changes in the probability distribution. In particular, $\bigtriangleup t$ fits neatly into the transition probability,
\begin{equation}
P(x^{\prime }|x,\Delta t)=\frac{1}{Z}\exp [-\sum_{n,a}(\frac{m_{n}}{2\eta \Delta t}%
\left( \,\Delta x_{n}^{a}-\langle \Delta x_{n}^{a}\rangle \right) \left(
\,\Delta x_{n}^{a}-\langle \Delta x_{n}^{a}\rangle \right) ]~,
\label{trans}
\end{equation}
as a measure of fluctuations, such that the state of knowledge of the positions of particles at a later time $t'$ is given by marginalizing over the previous position coordinates,
\begin{equation}
\rho(x'|t')=\int  P(x^{\prime }|x,\Delta t)\rho(x|t)\,dx,\label{step}
\end{equation}
where $\rho(x|t)\equiv \rho(x)$. Equation (\ref{step}) may be recast as a differential equation for $\rho(x|t)$ (an explication may be found in \cite{book}),
\begin{equation}
\partial _{t}\rho =-\partial _{A}\left( \rho v^{A}\right),   \label{FP b}
\end{equation}%
which may be recognized as the Focker-Planck equation because the current ``velocity" $v^{A}$ of the probability flow in configuration space is,
\begin{equation}
v^{A}=m^{AB}\partial _{B}\Phi \quad \mbox{where}\quad m^{AA^{\prime
}}=m_{n}^{-1}\delta ^{aa^{\prime }}\delta _{nn^{\prime }}\quad \mbox{and}%
\quad \Phi =\eta \phi -\eta \log \rho ^{1/2}~.  \label{curr}
\end{equation}
This form of probability updating, from maximum entropy considerations, leads to a diffusion equation of the $N$ particles. To derive QM, we need an additional mechanism for updating the probability of the positions of the particles. We impose $\Phi$ to be a dynamical variable that further updates $\rho$, and briefly, \cite{ED} finds QM to take the form of a non-dissipative diffusion. This is done by letting the evolution of $\rho $ and $\Phi $ be dynamically coupled (Hamilton's) equations,
\begin{equation}
\partial _{t}\rho =\frac{\delta \tilde{H}}{\delta \Phi }\quad \mbox{and}%
\quad \partial _{t}\Phi =-\frac{\delta \tilde{H}}{\delta \rho }~.
\label{HJ}
\end{equation}%
 The \textquotedblleft ensemble\ Hamiltonian\textquotedblright\ $\tilde{H}$
is chosen so that the first equation above reproduces (\ref{FP b}) (which in principle may always be done) in which case
the second equation in (\ref{HJ}) becomes a Hamilton-Jacobi equation. A
more complete specification of $\tilde{H}$ is,
\begin{equation}
\tilde{H}[\rho ,\Phi ]=\int dx\,\left[ \frac{1}{2}\rho m^{AB}\partial
_{A}\Phi \partial _{B}\Phi +\rho V+\xi m^{AB}\frac{1}{\rho }\partial
_{A}\rho \partial _{B}\rho \right] ~,  \label{Hamiltonian}
\end{equation}%
where the \textquotedblleft kinetic\textquotedblright\ $(d \Phi)^2$ terms 
reproduces (\ref{FP b}), and the potential terms, $V(x)$ and a \textquotedblleft
quantum\textquotedblright\ potential, are forcibly independent of $\Phi$ as they originate from integration constants in the solution for $\tilde{H}$ satisfying (\ref{FP b}). The form of the potential terms is suggested by information geometry in the presence of this inference framework \cite{ED}. Nothing prevents the combining of $\rho $ and $\Phi $ into a single complex function $%
\Psi \sim\rho ^{1/2}\exp (i\Phi /\hbar )$, and after some massaging which will not be reviewed here, ED reproduces the \emph{linear }Schr\"{o}dinger equation,%
\begin{equation}
i\hbar \frac{\partial \Psi }{\partial t}=-\sum_{n}\frac{\hbar ^{2}}{2m_{n}}%
\bigtriangledown^2_n\Psi +V\Psi ~,  \label{sch c}
\end{equation}
as an application of inference. At this point the standard Hilbert space formalism may be adopted to represent
the epistemic state $\Psi (x)$ as a vector, 
\begin{equation}
|\Psi \rangle =\int dx\,\Psi (x)|x\rangle \quad \mbox{with\quad }\Psi
(x)=\langle x|\Psi \rangle ~.
\end{equation}%
The expression of $\ket{\Psi}$ in another basis is regarded as a potentially convenient way of expressing position space wavefunctions.

A more general SE equation, which includes the presence of external nonzero electromagnetic vector potentials $\vec{A}$, may also be generated by ED \cite{ED}. This is done by applying an additional expectation value constraints  ${\bigtriangleup x^a}A_a(x_n)=\kappa''_n$ when maximizing (\ref{entropy}), similar in form to the drift potential from equation (\ref{kappa prime}). Spin is generated by positing the existence of a ``spin frame" field $\vec{s}(x)$ (motivated through geometric algebra), which again constrains the expected drift of the particles. The drift potential $\phi$, electromagnetic vector potential, and spin frame fields are introduced as epistemic parameters and update the position space distributions of quantum mechanical particles - they are only as ontic as the changes in probability distributions they generate, but probability distributions are epistemic in general. The Pauli equation for a single particle has been found using a spin frame in ED \cite{private}. The correct spin statistics for identical multi-particle states has yet to be generated from ED, so at this point we impose a symmetrization postulate \emph{ex post facto}, which of course is no better or worse than the standard quantum mechanical formalism. 
\subsection{Measurement in ED\label{measurement}}
 A natural question is, ``If position is the only definite quantity, how are other operators in Quantum Mechanics measured?". This question was originally addressed in \cite{Johnson} and more recently is addressed in \cite{Kevin} as well as how the notions of von Neuman, weak measurements, and Weak Values \cite{AAV, Duck, Dressel} fit into ED. This question is only addressed to the extent ``operators" and ``measurement" are defined within the Entropic Dynamics framework. Measurement is a two step process: the state of a system is first updated via a unitary and Schr\"{o}dinger evolution for the purpose of \emph{detection}, which is a Bayesian update made due to the presence of data. As ED is an application of inference and probability updating, measurement is simply tackled by applying the appropriate rules of inference. Because position is the only beable in ED, a state vector $\ket{\Psi}$ which is expanded in the basis $\ket{a}$, is a potentially convenient stand-in for position space wavefunctions,
\begin{eqnarray}
\ket{\Psi}=\sum_ac_a\ket{a}=\sum_ac_a\int\psi_a(x)\ket{x}\,dx=\int\Psi(x)\ket{x}\,dx.\label{13}
\end{eqnarray}
An operator such as $\hat{A}$ is an epistemic object, which does not require an ontic existence within the ED framework; however, their values may still be inferred. Operators, Weak Values, and eigenvalues are therefore a subset of a type of quantity we call the \emph{inferables} of the theory \cite{Kevin}. As it is the position of particles which formulate the ontic objects in ED, we make inferences about inferables by detecting positions.

To accomplish this, \cite{Johnson} introduces the concept of a unitary measurement device $\hat{U}_{A}|a_{i},t\rangle=|x_{i},t'\rangle$, which maps states $\ket{a}$ (the state we wish to infer) at time $t$ to a position on a screen $x_a$ at a later time $t'$. This allows for the inference of $\ket{a}$ (the eigenvectors of an operator $\hat{A}$) by making detections of $x_a$ at a later time. An example of such a unitary measurement device is a periodic crystal lattice or prism which diffracts ``momentum" states to position states. We have,
\begin{eqnarray}
\ket{\Psi',t'}=U_A\ket{\Psi,t}=\sum_a\ket{x_a,t'}\braket{a,t}{\Psi,t}=\sum_ac_a\ket{x_a,t'},\label{unitary}
\end{eqnarray}
such that $p(x_a|t')=p(a|t)=|c_a|^2$, that is, the particle may be detected at $x_a$ with probability $p(x_a|t')$ at a later time \emph{as if} it were earlier in the state $\ket{a}$. Inferences can then be made about the operator $\hat{A}=\sum_a\lambda_a\ket{a}\bra{a}$ where $\lambda_a$ are \emph{arbitrary} scalars. The actual detection of the location of a particle in a single experiment is facilitated with another detection device such as a photo-plate, CCD camera, or bubble chamber. In such instances, the probability of $x$ given a detection $D$ is given by,
\begin{eqnarray}
P'(x)=q(x|D)=\frac{q(x,D)}{q(D)}=\frac{p(x)q(D|x)}{q(D)},\label{detect}
\end{eqnarray}
where $q(D|x)$ is the likelihood function which accounts for the accuracy of the measurement device. In the present case, after a detection at $D$,
\begin{eqnarray}
P'(x_a|t')=q(x_a|D,t')=\frac{p(x_a|t')q(D|x_a,t')}{q(D)},
\end{eqnarray}
effectively ``collapses" the wavefunction, which is to say the final state of the system is known with certainty for sharply peaked likelihood functions in ED \cite{Kevin}. Similarly, if we want to infer the spin of a particle, we preform a von Neumann or a weak measurement to entangle the position and spin of the particle (via unitary evolution as one would with a Stern-Gerlach device) such that by detecting position, we may infer the spin in a similar fashion.
\subsection{Remark}
Quantum Mechanics has been derived as a peculiar application of epistemic probability updating when the ontic elements of interest are the positions of particles. No further interpretation of Quantum Mechanics in ED is needed. The wavefunction is found to be a useful epistemic quantity for calculating probability distributions, which represent the state of knowledge of a system. Other quantum mechanical objects, like operators and Hilbert spaces, play a supporting role.

Concepts in ED are naturally communicated in the the language of probability. The language generated by the Copenhagen interpretation of Quantum Mechanics clashes somewhat with the language of probability; for instance, the notion of an ``observable" makes little sense when nontraditional Hermitian operators are considered, i.e. does one ever really claim to ``observe" $\hat{p}^n$, $\hat{\rho}$, or one of its eigenvalues? In truth these quantities are inferred through the measurement process no differently than the average energy of a statistical system is inferred by measuring the height of mercury on a thermometer. What was observed in this scenario is the height of the mercury and what is inferred is the temperature and the energy of the system. The word ``observable" really looses meaning when one considers ``measuring" the Weak Value of an operator $A_W=\frac{\bra{\Psi'}\hat{A}\ket{\Psi}}{\braket{\Psi'}{\Psi}}$, which in general is a complex numbers that may lie outside the eigenvalue spectrum of the operator $\hat{A}$ \cite{AAV, Duck, Dressel}. In ED, Weak Values are simply categorized as potentially interesting inferables of the theory as they are inferred from pointer variable (positions in ED) detections \cite{Kevin}. The language used in contextuality proofs does not naturally coincide with the language of probability and inference, which is touched upon later.

\section{\label{sec:level1}$\psi$-epistemic?}

In the previous section we claimed that $\psi$ is an epistemic object which represents our current knowledge of the system in question. This immediately runs into conflict with the $\psi$-epistemic \emph{no-go} theorem from \cite{Pusey}; however, there is no issue. An excellent review of the $\psi$-epistemic/ontic dichotomy is presented in \cite{Leifer} and ED would be categorized as a realist (or partial realist) $\psi$-epistemic model. The first assumption in \cite{Pusey} is (paraphrased) 1) is that a $\psi$-epistemic state has physical values upon which inferences may be made. ED agrees with this assumption whole heartedly, and the variables which are ``physical" in ED are alone the definite yet unknown positions of particles.  The second assumption (verbatim) 2) is that ``systems which are prepared independently (a) have independent physical states (b)". 

The second assumption requires further investigation: first (a)  - what does it mean for a system to be prepared independently and second (b) - what does it mean for a system to have independent physical states? The definition of independence in (a) seems to be saturated by the definition of independence in probability theory, namely that if two systems are prepared independently then their joint probability distribution is factorisable into independent probability distributions, $p(x_1,x_2)=p_1(x_1)p_2(x_2)$, and therefore there are no correlations between $x_1$ and $x_2$ at that time (however evolution may later induce correlations). The quantum mechanical analog is that these two states are non-entangled product states. The definition of independent physical states in (b) is rather unclear from the outset but later is defined quantitatively by,
\begin{eqnarray}
D(\mu_0,\mu_1)=\frac{1}{2}\int|\mu_0(\lambda)-\mu_1(\lambda)|d\lambda,
\end{eqnarray}
or equivalently in our notation,
\begin{eqnarray}
D[p_1,p_2]=\frac{1}{2}\int|p_1(x_1)-p_2(x_2)|\delta(x_1-x_2)dx_1dx_2,
\end{eqnarray}
 such that if $D=1$ then $p_1$ and $p_2$ are completely disjoint and thus occupy (in their words) ``independent physical states". It is easy to see now how the definition of independent in (a) differs from the definition of independent in (b), in-fact the (a) and (b) definitions of independent clash in assumption 2) for any independent joint probability $p(x_1,x_2)=p_1(x_1)p_2(x_2)$ (a) which are not entirely disjoint (i.e. ``physically" independent (b)) -- that is, if $p_1(x_1)$ and $p_2(x_2)$ overlap in $\mathcal{X}$. This regularly occurs in noninteracting multiparticle states in Quantum and Statistical Mechanics ( can be obtained by marginalizing over the momentum of a phase-space probability distribution $\rho(x,p)$). In ED, the ``physicality" of particle positions is independent of the state of knowledge at hand. This is because probability is not a measure of physicality (or onticity) but rather as a degree of belief or plausibility \cite{book, Jaynesbook, Cox} that the proposition ``the particle is located at $x$" is true. As the leading assumptions of what entails a $\psi$-epistemic state differ, the $\psi$-epistemic \emph{no-go} theorem does not apply, which is admitted as a possible exemption to their \emph{no-go} theorem in the conclusion of \cite{Pusey}. We are therefore justified in treating $\psi$ epistemically. 
\section{\label{sec:level1}Hidden Variables, Realism, and Non-locality}
The subject of hidden variables, realism, and non-locality in ED has been touched upon in \cite{book,Johnson} and it will be further explored here. In Bell's landmark paper \cite{BellEPR}, he found a contradiction between QM and hidden variable theories which claimed local realism. It was accomplished by considering a hidden variable $\lambda$, which if known, would give the outcome of an experiment (an eigenvalue of an operator) with certainty $a_0=A(\lambda=\lambda_0)$. By integrating over the probability of a hidden variable,
\begin{eqnarray}
\expt{A(\lambda)B(\lambda)}=\int p(\lambda) A(\lambda)B(\lambda)\, d\lambda,
\end{eqnarray}
he showed that such expectation values do not always agree with the expectations values of QM, for general $p(\lambda)$. 

In ED there is no such hidden variable. The particle dynamics are non-deterministic as can be seen by the Brownian like paths particles take due to the form of the transition probability $P(x'|x)$ in (\ref{trans}), or after energy conservation, that the particles are undergoing a non-dissipative diffusion. Even if the initial conditions of particle positions are known exactly (or with near perfect precision), $\rho(x)\sim\delta(x-x_0)$, equation (\ref{step}) is inevitably nondeterministic for time steps $\Delta t$. Because the Brownian paths of particles are non-differentiable, other equi-temporal quantities (e.g. momentum or energy) are simultaneously indefinite, which is another argument against position being a hidden variable.  The process which is deterministic in ED is the evolution of the probability distribution as it follows the HJ-like equations from (\ref{HJ}) given the appropriate constraints, boundary, and initial conditions are known. The drift gradient $ \frac{\partial \phi(x)}{\partial x^a_n}\sim \Phi$ updates the probability distribution of particle locations rather than guiding each particle at every point, again seen by inspecting (\ref{HJ}).  The solution is to realize that $\Phi(x)$ is an epistemic parameter that is coupled to $\rho(x)$ for each $x$ in a complicated way through the HJ-like equations (\ref{HJ}).  The nonlocal nature of probability as a means for quantifying knowledge (of the future, past, or present) accounts for the nonlocal behavior of QM in ED. As any collapse is an epistemic change in the system, each observer assigns distributions which coincide with their current state of knowledge of the system.

 \section{\label{sec:level1}Bell-KS type Theorems}

The Bell-KS theorem sheds light on the incompatibility of hidden variable theories and Quantum Mechanics \cite{Bell, KS}. Years later Mermin demonstrated what is considered to be the simplest expression of what is usually an algebra and geometry intensive Bell-KS theorem \cite{Mermin}. Bell-KS proofs have been generalized to the $N$-qubit Pauli group \cite{Cai} and \cite{Cabellocont} gives a Bell-KS proof using continuous position and momentum observables. In \cite{Cai}, they give a simple algorithm to convert observable based KS proofs to a large number of projector based KS proofs, so we will focus on the simpler observable based proofs.

The class of hidden variable theories the Bell-KS theorem excludes have the following reasonable conditions: The value of an operator is \emph{definite} yet \emph{unknown} such that we may assign it a preexisting value (its eigenvalue) called its valuation \cite{Mermin, Cai, Ballentine}. The valuation of an operator $\hat{A}$ at any time is then,
\begin{eqnarray}
v(\hat{A})=\bra{a}\hat{A}\ket{a}=a.\label{val}
\end{eqnarray}  
It is also assumed that functional relationships between operators $f(\hat{A},\hat{B},\hat{C},...)=0$ hold throughout the valuation process,
\begin{eqnarray}
v(f(\hat{A},\hat{B},\hat{C},...))=f(v(\hat{A}),v(\hat{B}),v(\hat{C}),...)=0.\label{func}
\end{eqnarray}
It should be noted that the considered operators must commute $v(\hat{A}\hat{B})=v(\hat{A})v(\hat{B})=v(\hat{B}\hat{A})$ when taking valuations for (\ref{func}) to hold. Mermin demonstrates the contradiction of equations (\ref{val}) and (\ref{func}) with Quantum Mechanics by considering what is now know as the Peres-Mermin Square: 
\begin{center}
\begin{tabular}{ |c|c|c| }
\hline
 ZI & IX & ZX \\ 
\hline
 IZ & XI & XZ \\  
\hline
 ZZ & XX & YY\\
\hline
\end{tabular}
.
\end{center}
Each table entry is an observable from the 2-qubit Pauli group consisting of a joint eigenbasis consisting of 4 eigenvectors. As a notational convenience we will omit tensor products when there is no room for confusion and let $X=\sigma_x$ such that an arbitrary table entry $XZIY$ represents $\sigma_x^{(1)}\otimes\sigma_z^{(2)}\otimes I^{(3)}\otimes\sigma_y^{(4)}$, following the notational structure in \cite{Cai}.  The product of the operators along a given row or column is the rank 4 identity $I(4)=II$ (in this notation) with the exception of the last row, which is  $-II$. Consider the valuation of the standard matrix product of the elements of the first row,
\begin{eqnarray}
v(ZI \cdot IX \cdot ZX)=v(II)=1.
\end{eqnarray}
Supposing (\ref{func}) is true then
\begin{eqnarray}
v(ZI \cdot IX \cdot ZX)=v(ZI)v(IX)v(ZX)=1.
\end{eqnarray}
The valuation of the $ij$th element $A^{ij}$ in the table is $v(A^{ij})=\pm 1$, and therefore (\ref{func}) imposes a constraint on the individual valuations $v(ZI)v(IX)v(ZX)=1$, which is only satisfied if either 0 or 2 of the valuations are $-1$. This cuts the universe of discourse from $2^3=8$ possibilities down to $4$. Let $A^{i\odot}$ be the product of the operators in the $i$th row and $A^{\odot j}$  the product of the operators in the $j$th column such that above $A^{1\odot}=$$ZI\,\cdot\,$$IX\,\cdot\,$$ZX$ is the standard matrix product between the listed operators. Mermin showed his square indeed leads to a contradiction when considering the product of the row and column valuations,
\begin{eqnarray}
\prod_{i}v(A^{i\odot})v(A^{\odot i})=v(II)^5v(-II)=-1,\label{agree}
\end{eqnarray}
whereas applying (\ref{func}) to each row and column, $v(A^{i\odot})=\prod_jv(A^{ij})$, gives,
\begin{eqnarray}
\prod_{i}v(A^{i\odot})v(A^{\odot i})\rightarrow \prod_i\prod_jv(A^{ij})^2=1\label{contradiction},
\end{eqnarray}
which is a contradiction. This is due to the fact that not all of the elements in Mermin's square commute and therefore all observables cannot be assigned definite eigenvalues. Quantum mechanical formalism and experiment agrees with (\ref{agree}) and not with (\ref{contradiction}) and thus (\ref{func}) must be thrown out. Bell makes a point that it may be overconstraining for the valuation to produce identical values when different sets of commuting observables are being considered, just to refute it by noting that a space-like separated observer could change which set of commuting observables he/she wishes to measure mid-flight. A hidden variable theory would then have to explain this nonlocal change in the valuation meaning that the Bell-KS theory only refutes local hidden variables theories.

\subsection{\label{sec:level1}Interpreting the Contradiction: Contextuality}

The standard interpretation of the contradiction by Bell, Kochen, Specker, Mermin and others is that quantum mechanical observables are \emph{contextual}, meaning that the operator's ``aspect", ``character", or ``value" depend on the remaining set of commuting observables under which it is considered.  Any observable which does not depend on the remaining set of commuting observables in this way is called \emph{noncontextual}, which, for example, are the individual observables $v(A^{ij})$ from the Mermin square and (\ref{contradiction}). 

In more recent years the interpretation of the Bell-KS theorem, which in principle would rule out all local hidden variable theories obeying (\ref{val}) and (\ref{func}), has been under scrutiny, in essence, for having a more restrictive interpretation than the theorem merits. The work by \cite{Meyer, Kent, Clifton} opens a loophole due to the impracticality of infinite measurement precision, and thus the Bell-KS theorem is ``nullified" in their language. Appleby (and others) find the ``nullified" critique to be too harsh of a criticism \cite{Appleby}. De Ronde \cite{deRonde} points out that epistemic and ontic contextualty are consistently being scrambled into a omelet when perhaps the yoke and egg whites should be cooked separately.  He defines ontic contextuality as the formal algebraic inconsistency of the operator and valuation formalism of Quantum Mechanics within the Bell-KS theorem -- having nothing to do with measurement. The epistemic counterpart is more aligned with the principles of Bohr in that Quantum Mechanics involves an interaction between system and measurement apparatus whose outcomes are inevitably communicated in classical terms -- the context is given by the measurement device. The difference is subtle but, as noted, ontic contextuality is defined to be independent of the differing interpretations of quantum mechanics where epistemic contextuality need not be. Our treatment of contextuality does separate in this fashion; however, de Ronde's usage of the word ``ontic" refers to the quantum formalism, whereas our usage only refers to ontic particle positions in ED.

\subsection{Critiques\label{crit}}
As we know, the assumptions (\ref{val}) and (\ref{func}) lead to contradictions. Inevitably (\ref{val}) and (\ref{func}) will be illogical on several levels, some of which are discussed below. The main critique we present is, how do we know that the valuation of an observable $v(\hat{A})$ accurately represents the notion of  \emph{definite, preexisting values} of an operator, that would be obtained if a measurement is carried out?  The alleged strength of the Bell-KS theorem is that analysis has been done independent of the particular state $\ket{\Psi}$ and thus it should hold for all $\ket{\Psi}$ in general. This is troubling for a number of reasons, the first being that a particular $\ket{\Psi}$ may not have components along every eigenvector of an operator $\hat{A}$, in which case a zero probability event could be assigned a definite existence, and one would never know because $\ket{\Psi}$, which all of the observables in question pertain to, has not been specified. This issue here is an interplay between the ontic and epistemic contextuality given by de Ronde, because only sensible valuations may be given if the state of the system is known -- in general the density matrix $\hat{\rho}$. 

If the valuation process is to be applicable to arbitrary ``observables" independent of the state at hand, then one runs into another logical inconsistency when attempting to apply valuations to a density matrix, $\hat{\rho}$, because it represents the probabalistic state of a system. It makes little sense to have different sets of commuting observables $\{\hat{\rho}_{1},\hat{\rho}_{2},\hat{\rho}_{3}...\}$ which are required to span to the same Hilbert space as the state in question $\ket{\Psi}$ (or $\hat{\rho}$). Furthermore, the valuation of a density matrix $\hat{\rho}=\sum_ip_i\ket{i}\bra{i}$ gives one of its eigenvalues, $p_i$, which are probabilities themselves and are never \emph{directly} observed, but are usually inferred from the frequency of a large number of independent trials. One cannot possibly claim that a system is ontically expressing a definite preexisting probability value $p_i$. Probability by its nature is a measure of the indeterminance of a state $\ket{i}$ rather than a value (physically) carried by the state $\ket{i}$  -- which is as epistemic as it gets! If Alice knowingly prepares one system and Bob does not know which system Alice has prepared, then it is clear that $p_i$'s cannot have a definite existence because both Alice and Bob disagree about said values over the same single ``ontic" system of interest. Furthermore, when the a measurement is made to determine the state, the probability value updates (the eigenvalue changes) and in this sense the assignment of an eigenvalue $\hat{\rho}$ through valuation represents nothing physical about the state of the system's \emph{definite, preexisting values} that would in principle be obtained if a measurement was carried out. In this sense, the eigenvalues of operators in general do not represent  \emph{definite, preexisting} (noncontextual) \emph{values} of an operator.

There are some inferences that can be made using probability theory (and thus ED) that formally cannot be made using the standard tools of quantum mechanics. For instance, one of the motivating arguments for the use of valuation functions in \cite{KS} is that the value of an operator $\hat{A}$ and its square $\hat{B}=\hat{A}^2$ clearly have highly correlated outcomes that should be represented though a joint probability distribution $P(a,b)$ (implied for a single experiment), which is incorrectly given as $P(a,b)=p(a)p(b)$. If a measurement of $\hat{A}$ is preformed such that the probability of a particular outcome is $p(a)=|\braket{a}{\Psi}|^2$, but the eigenvalues of $\hat{B}$ are related to the eigenvalues of $\hat{A}$ by a known condition $b=f(a)$, the desired joint probability distribution which relates the outcome of $a$ to the outcome of $b$ is,
\begin{eqnarray}
P(a,b)=p(a)p(b|a)=p(a)\delta(b-f(a)),
\end{eqnarray}
 where $\delta(b-f(a))$ is the Dirac/Kronecker delta function if $\hat{A}$ has a continuous/discrete spectrum, which is obtained through standard probability theory. The probability of finding a value $b$ without any knowledge of the particular $a$ is obtained by marginalizing over $a$,
\begin{eqnarray}
p(b)=\sum_{a}P(a,b)=\sum_{a}p(a)\delta(b-f(a)).
\end{eqnarray}
The pedagogical case where every eigenvalue $a$ has a negative counterpart (except the possible $a=0$) and $\hat{B}=\hat{A}^2$, such that $f(a)=a^2$ leads to
\begin{eqnarray}
p(b=a^2)=\sum_{a}p(a)\delta(b-a^2)=p(a)+p(-a),
\end{eqnarray}
the probability of $b$ takes the value $a^2$ is the sum of the probabilities of $a$ and $-a$. Trying to make this inference from QM would require defining states like $\ket{\Psi}\sim\sum_{ab}c_a\delta_{b,a^2}\ket{a}\otimes\ket{b}=\sum_ac_a\ket{a}\otimes\ket{a^2}$ and using Born's Rule; however, vectors of that form are redundant for a single particle as both $\hat{A}$ and $\hat{B}$ both share the eigenbasis $\{\ket{a}\}$ which spans a single Hilbert space $\ket{a}\otimes\ket{a^2}\rightarrow \ket{a}$ rather than a tensor product of two, and for that reason, vectors of this type are thrown out. 

Due to these critiques, and that in ED one may infer eigenvalues from position detections, it is difficult to know what precisely a valuation procedure represents meta-physically (linguistically), besides the simple choice of a matrix element. As discussed, the valuation of an operator may not always represent an ontic value of an observable, and therefore we suggest relaxing this notion and replacing it by the more general statement, ``The valuation of an operator (or set of operators) represents a quantity that in principle may be inferred", or in the language of \cite{Kevin}, ``The valuation of an observable is an \emph{inferable} of the theory".

\subsection{\label{sec:level1}Hybrid-contextual theories}

It should be noted that in Entropic Dynamics, the idea of valuation is very unnatural. An inference based theory allows us to state, quantify, and represent how much we do \emph{not} know about the state of a system through a probability distribution, upon which we use the rules of inference and probability updating to determine what we do. Noted earlier in Section \ref{measurement}, the measurement procedure in ED allows for the inference of $\hat{A}=\sum_a\lambda_a\ket{a}\bra{a}$ where $\lambda_a$ are arbitrary scalars, by making detections of position at a later time. Eigenvalues themselves are an afterthought of the inference process that are epistemically inferable parameters by the changes they make to a probability distribution.

Strictly speaking, the Bell-KS theorem discards realist theories in which all of the considered operators are treated ontically through their valuation. This leaves open the possibility for a hybrid-contextual theory in which only a subset of commuting observables are \emph{definite yet unknown}, or noncontextual, while other variables (or sets of commuting observables) are contextual. To date the only theory of Quantum Mechanics known to the author that seems to fit this description precisely is Entropic Dynamics \cite{ED}.  


The only operators required to undergo valuation in ED are the $3N$-particle position coordinates with their corresponding $3N$ operators $\hat{X}^{(n)}$. In the language of valuations, we would have,
\begin{eqnarray}
v_x(\hat{x}^{(n)}_i)\equiv \bra{x_i^{(n)}}\hat{x}^{(n)}_i\ket{x_i^{(n)}}=x^{(n)}_i,
\end{eqnarray}
for a particular coordinate $x_i^{(n)}$.
Position operators trivially obey (\ref{func}),
\begin{eqnarray}
v_x(f(\hat{x}^{(n)}_i,\hat{x}^{(m)}_j,...))=f(v(\hat{x}^{(n)}_i),v(\hat{x}^{(m)}_j),...)=0,
\end{eqnarray}
for any function $f$, because all position operators mutually commute.  No parity contradiction in the sense of \cite{Mermin,Cai} can be reached because all of the operators requiring valuation mutually commute. The Bell-KS proofs are proofs by contradiction.  This means a set of counter examples has been found which rule out the general applicability of assigning definite yet unknown values to all operators all the time; however, as seen above, there are instances in which there is no contraction and the assignment of definite yet unknown values in this instance is consequently \emph{not} ruled out.

Operators other than position, $A^{ij}$, need not be noncontextual in ED as they are considered to be epistemic in nature. In this case, one should not claim $A^{ij}$, one of its eigenvalue $a^{ij}$, or a state $\ket{a^{ij}}=\int dx \,\psi_{a^{ij}}(x)\ket{x}$, to have a definite existence outside of characterizing our knowledge of the definite yet unknown positions of particles $x$. That being said, when one can expand $A^{ij}$ in the position basis, we find that the $A^{ij}$ are naturally contextual -- although in principle this is unwarranted in ED as no valuation is required.

 As it is the position that is definite, the valuation of the operator $A^{ij}$, before measurement (where $\ket{x}=\ket{x_1}\otimes\ket{x_2}...\otimes\ket{x_N}$ for $N$ particles), is one of the diagonal matrix elements in the $x$ basis,
\begin{eqnarray}
v_x(A^{ij})\rightarrow \bra{x_0}A^{ij}\ket{x_0}=\bra{x_0}\sum_n\ket{a_n^{ij}}a_n^{ij}\braket{a_n^{ij}}{x_0}=\sum_na_n^{ij}|\braket{x_0}{a_n^{ij}}|^2\neq a_n^{ij},
\end{eqnarray}
where in this case it is supposed that the definite yet unknown value of $x$ is $x_0$. This is obviously not one of the eigenvalues or ``observables" of $A^{ij}$, but in ED $A^{ij}$ is an inferable and so is $v_x(A^{ij})$. The position space valuation $v_x(A^{ij})$ is some real number which in principle may be assigned to any position coordinate.  In general, parity type proofs of the Bell-KS theorem require  $A^{ij}$ to be simultaneously part of an even number of sets of commuting observables \cite{Cai}. This means an operator $A^{ij}$ is simultaneously diagonalized in (at-least two) different basis,
\begin{eqnarray}
A^{ij}=\sum_n\ket{a_n^{i\odot}}a_n^{ij}\bra{a_n^{i\odot}}=\sum_n\ket{a_n^{\odot j}}a_n^{ij}\bra{a_n^{\odot j}},
\end{eqnarray}
where, for example in the Peres-Mermin square, $\ket{a_n^{i\odot}}$ refers to the eigenvectors of the commuting set of variables from the $i$th row and $\ket{a_n^{\odot j}}$ refers to the eigenvectors of the commuting set of variables from the $j$th column. The largest number of distinct sets of eigenvectors is equal to the number of sets of commuting observables in the Bell-KS proof. Using this notation we may denote the product of the operators in a commuting set by,
\begin{eqnarray}
A^{i\odot}=\sum_n\ket{a_n^{i\odot}}a_n^{i1}a_n^{i2}...a_n^{iN}\bra{a_n^{i\odot}}=\sum_n\ket{a_n^{i\odot}}a_n^{i\odot}\bra{a_n^{i\odot}},
\end{eqnarray}
where $N$ is the number of operators in the commuting set of observables. In general, the application of (\ref{func}) to the position valuations of $\{A^{ij}\}$ will not hold,
\begin{eqnarray}
v_x(A^{i\odot})\rightarrow \prod_jv_x(A^{ij}),
\end{eqnarray}
because it would require,
\begin{eqnarray}
\sum_n|\braket{x_0}{a_n^{i\odot}}|^2a_n^{i\odot}\rightarrow \prod_j\Big(\sum_na_n^{ij}|\braket{x_0}{a_n^{i\odot}}|^2\Big)_j,
\end{eqnarray}
which is potentially equal, but in the vast majority of cases is not. The lack of equality can be seen if one considers three commuting momentum observables $\hat{p}_1\otimes\hat{1}_2$, $\hat{1}_1\otimes\hat{p}_2$, and $\hat{p}_1\otimes\hat{p}_2$ with $\{\ket{a_n^{i\odot}}\}=\{\ket{p_1,p_2}\}$ -- the LHS diverges while  the RHS is zero because it involves products of odd integrals. This poses no issue in ED because $A^{i\odot}$ or the individual $A^{ij}$ need only exist epistemically, so their valuations (matrix elements) need not agree - the product of matrix elements need not be the matrix element of the product so imposing equality is nonsensical. Furthermore (if above was not enough), contextuality is preserved among non-position observables (for noncontextual position) as can be seen when (\ref{func}) is applied to the product of all of the commuting sets of observables, 
\begin{eqnarray}
\prod_{i}v_x(A^{i\odot})v(A^{\odot i})\rightarrow \prod_i\prod_jv_x(A^{ij})^2\geq0,
\end{eqnarray}
for situations when the LHS is less than zero or it is simply not equal to the RHS. This calculation shows that definite (noncontextual) positions before measurement do not imply definite (noncontextual) $A^{ij}$, and therefore, we are justified in treating the operators $A^{ij}$ contextually - which means we should not apply valuations to them, or if we do, we should not expect (\ref{func}) to hold. Spin in ED is not required to be noncontextual so we don't need to apply valuations to them either.  The current form of ED would potentially be ruled out if $A^{ij}$ were noncontextual under position valuations in general - but this is not the case as the matrix elements are simply epistemic inferables. 

Because all position operators always mutually commute with one another and therefore are all simultaneously diagonalizable in the same set of position eigenvectors (i.e. $\ket{x}=\ket{x^{i\odot}}=\ket{x^{\odot j}}=\ket{x^{\odot\odot}}$), they may be treated noncontextually together. If an operator is a product of contextual and noncontextual operators, it remains contextual because applying position space valuations on the noncontextual operators leaves the contextual operators contextual. This can be seen by applying position space valuations to the continuous operators defined in \cite{Cabellocont} ($|\expt{S_{QM}}|=6$). As noted in the critiques, the valuation of an operator may not always express the definite yet unknown values of an observable --  it may be best to relax this notion such that the valuation of an operator represents a quantity that in principle may be inferred, an inferable, in general.

A question of interest is, how, if everything is to be measured or inferred using a (non-contextual) position basis (Section \ref{measurement} and \cite{Johnson,Kevin}), is the contextual nature of a set of contextual operators $\hat{A}^{ij}\in \mathcal{A}$ non-contradictory? This question is especially tricky because it mixes the epistemic and ontic palates of contextuality in the sense of \cite{deRonde}, who, as well as \cite{Cabello}, quote Mermin , ``the whole point of an experimental test of KS [theorem] misses the point.". That being said, the contexuality of the operators $A^{ij}$ is simply expressed through the lack of commutativity between sets of commuting observables, we do not need to do position valuations $\{A^{ij}\}$ to make inferences about the states. ED perhaps sheds some light onto Mermin's statement about the lack of an experimental test. 

Suppose Alice prepares a two particle system and sends it to Bob who has a compound unitary measurement device (\ref{unitary}) for each set of commuting observables (each row and column) of the Mermin square (for simplicity), but really this is applicable to any construction of sets of commuting observables. Because Bob can only measure one row or column for a given pair of particles sent from Alice, him choosing a measurement device from the $i$th row or column means he has chosen and applied the unitary measurement device $\hat{U}_{A^{i\odot}}$ to the incoming state and mapped it to position coordinates from which inferences can be made. That is, the physical application of $\hat{U}_{A^{i\odot}}$ picks, $P^i$, the $i$th row,

\begin{center}
$P^i \,*\,$
\begin{tabular}{ |c|c|c| }
\hline
 $A^{11}$ & $A^{12}$ & $A^{13}$ \\ 
\hline
 $A^{21}$& $A^{22}$ & $A^{23}$\\  
\hline
 $A^{31}$& $A^{32}$ & $A^{33}$\\
\hline
\end{tabular}
$\longrightarrow \hat{U}_{A^{i\odot}}$ \begin{tabular}{ |c|c|c| }
\hline
 $A^{i1}$ & $A^{i2}$ & $A^{i3}$ \\ 
\hline
 \end{tabular}
$\longrightarrow$
\begin{tabular}{ |c|c|c| }
\hline
 $x^{i1}$ & $x^{i2}$ & $x^{i3}$ \\ 
\hline
 \end{tabular}
,
\end{center}
and at a later time one may apply valuation(s) to the associated positions operators if one wishes, because the operator is position based (at that later time) $A^{ij}\rightarrow \sum_n\ket{x^{ij}_n}a^{ij}_n\bra{x^{ij}_n}$. The positions may be detected and the associated commuting set of $A^{ij}$ may be inferred. 


This process resembles the epistemic notion of contextuality presented in \cite{deRonde}.   The operators $A^{ij}$ are treated contextually -- the position space valuation may be applied after the set of commuting observables has been chosen by the unitary measurement device.  As only one column or row may be picked at a time by Bob, who seeks to measure a set of commuting observables, the quantum mechanical expectation values match that as read by Bob (and are therefore in the form of (\ref{agree})). Alice, being in the dark, does not know which row Bob will pick and is free to assign a probability Bob picks the $i$th row or column, and after learning the chosen row or column may she update her probability accordingly.

\section{\label{sec:level1}Discussion}

The most natural inferential tool in ED is probability. The critiques given in Section \ref{crit} are further motivation for the use of probability to make rational inferences, while the interpretation of the valuation functions, which inevitably lead to contradictions in the Bell-KS theorem, are suspicious by several accounts.  There, reason was given for the need of a more general interpretation of the valuation of an operator, which was stated, ``The valuation of an operator (or set of operators) represents a quantity that in principle may be inferred". Because the probability of a state is only defined in terms of its set of commuting observables, and because there is no way to generate a unique joint probability distributions among non-commuting observables \cite{Cohen}, a rational discussion on the potential simultaneous onticity between non-commuting observables is not possible -- luckily ED formulates QM by assuming the position of particles to be the only ontic variables.

ED is in a unique position among foundational theories of QM because QM was derived by applying standard probability techniques to a system having ontic positions of particles. This naturally classifies the ontic and epistemic elements of QM and provides a clean cut interpretation of QM such that physical and conceptual problems in QM may be handled rigorously (as it has  in \cite{Johnson, Kevin} and other recent articles). The hope is that a full treatment of spin in ED will provide better notions of the symmetrization postulate and the Pauli exclusion principle, but this problem has yet to be tackled in full. At this point more work can be done to make sure that ED is able to reproduce all of the known results of QM and to hopefully shine some light on the many interpretational paradoxes, \emph{no-go} theorems, and problems that surround QM. The end goal of ED is to show inferential origins of physical laws. This generates new interpretations and directions for old ideas and hopefully ED will generate some new physics as well.

 This paper shows the sense in which a foundational QM theory may be hybrid-contextual, i.e. one set of commuting operators is noncontextual (ontic) while all others are contextual, and still obey the Bell-KS theorem. In ED this occurs because contextual operators are not required to have definite ontological existence outside of their characterization of the state of knowledge of the noncontextual operators. A loose guide for a theory to be hybrid-contextual, and also agree with QM, is that its ontic variables are treated noncontextually while its epistemic variables are treated contextually. The values of interest associated with contextual operators (energy, momentum, and spin) are inferred by measurement of noncontextual observables (position in ED). 


\section{Acknowledgments:}
I would like to thank everyone in the information physics group at UAlbany, especially Ariel Caticha, Nick Carrara, Selman Ipek, and Tony Gai  who have encouraged me throughout the writing of this paper. I would also like to thank Mordecai Waegell and Christian de Ronde for our discussions about quantum mechanics and physics in general.

\bibliography{bib}



\end{document}